\documentclass{PoS}
\usepackage{amsmath}

\let\ifpdf\relax
\newcommand{\U}{{\rm U\,}}
\newcommand{\W}{{\rm W\,}}
\newcommand{\tr}{{\rm tr\,}}
\newcommand{\be}{\begin{eqnarray}}
\newcommand{\ee}{\end{eqnarray}}
\newcommand{\eins}{\leavevmode\hbox{\small1\kern-3.8pt\normalsize1}}

\title{The Effect of the Low Energy Constants on the Spectral Properties of the Wilson Dirac Operator}

\ShortTitle{Spectral Properties of the Wilson Dirac Operator}

\author{Mario Kieburg\\
        Fakult\"at f\"ur Physik, Postfach 100131, 33501 Bielefeld, Germany\\
        E-mail: \email{mkieburg@physik.uni-bielefeld.de}}

\author{Jacobus J. M. Verbaarschot\\
        Department of Physics and Astronomy,  Stony Brook University, NY 11794-3800, USA\\
        E-mail: \email{jacobus.verbaarschot@stonybrook.edu}}

\author{\speaker{Savvas Zafeiropoulos}\\
        Department of Physics and Astronomy,  Stony Brook University, NY 11794-3800, USA and\\
        Laboratoire de Physique Corpusculaire, Universit\'e Blaise Pascal, CNRS/IN2P3 
63177 Aubi\`ere Cedex, France\\
        E-mail: \email{zafeiropoulos@clermont.in2p3.fr}}

\abstract{We summarize recent analytical results obtained 
for lattice artifacts of the non-Hermitian Wilson Dirac operator. 
Hereby we discuss the effect of all three low energy constants. 
In particular we study the limit of small lattice spacing and also consider
the regime of large lattice spacing 
which is closely related to the mean field limit. 
Thereby we extract simple relations between measurable quantities like the average number of additional real modes 
and the low energy constants. These relations may improve the fitting for the low energy constants.}

\FullConference{31st International Symposium 
on Lattice Field Theory LATTICE 2013\\
                 July 29 - August 3, 2013\\
                 Mainz, Germany}

\begin{document}

\section{Introduction}

\label{intro}

One of the most widely used fermion discretizations is 
that of Wilson fermions. The increase of computational power 
and the improvement of algorithms has allowed for 
simulations in the deep chiral regime.
To extrapolate the lattice data to the continuum limit 
it is crucial to understand the lattice discretization effects 
as well as the interplay between finite lattice spacing 
and chiral symmetry breaking.
This problem was attacked in numerous ways. 
There has been a large ongoing effort to determine the 
Low Energy Constants (LECs) of Wilson chiral pertubation theory
\cite{SS} 
both from an analytical point  of view 
\cite{Damgaard:2010cz,Akemann:2010zp,NS,Kieburg,Kieburg:2011uf,Kieburg:2012fw,Kieburg:2013xta} 
as well as numerically 
\cite{Baron,Michael,757044,DWW,BBS,DHS,Herdoiza:2013sla} 
in order to elucidate the new phases that one encounters at finite 
values of the lattice spacing. Those discretization artifacts have 
also been studied for QCD-like theories like two-color QCD and QCD with 
adjoint fermions \cite{Kieburg:2013tca}.

In our analysis we incorporate the effect of the three  
LECs at order $a^2$. As a starting point we employ Wilson random matrix theory (RMT) for the non-Hermitian Wilson Dirac 
operator originally proposed in \cite{Damgaard:2010cz}. 
We refer the reader to \cite{KimLat12} for a review of 
the recent developments of RMT for the Wilson Dirac operator. This 
proceeding is a summary of the main results of 
Ref.~\cite{Kieburg:2013xta}. Readers who are interested in the 
derivation of those results are referred to this work and related
 mathematical developments 
published in \cite{Kieburg}.

The contents of this proceeding is as follows. 
In Sec.~\ref{effects} we briefly introduce the RMT and describe 
the effect of $W_6$ and $W_7$ on the spectrum of the 
Wilson Dirac operator by switching off $W_8$. 
In Sec.~\ref{densities} we summarize the effects of all three LECs 
on the level densities and discuss the limits of large and small 
lattice spacing. Thereby we provide comparisons of the analytical 
results with Monte-Carlo simulations of Wilson RMT. 
Moreover we propose some observables accessible by 
lattice simulations to extract the LECs.

\section{The effects of $W_6$ and $W_7$}\label{effects}

The introduction of the Wilson term in the lattice 
discretization of the Dirac operator explicitly breaks chiral symmetry. 
As a result the low energy effective theory for QCD will be 
affected and consequently new terms have to be incorporated 
in order to describe this effect. 
 Denoting the quark masses by $m$, the spacetime volume by $V$ 
and the chiral condensate by $\Sigma$, the effective partition function 
for $N_{\rm f}$ fermionic flavors reads
\begin{eqnarray}
 Z^\nu_{N_{\rm f}}(m)&=&\int\limits_{\U(N_{\rm f})}d\mu(U)\exp\left[\frac{\Sigma V}{2}\tr m(U+U^{-1})-a^2 VW_6\tr^2( U+U^{-1})\right]\nonumber\\
 &&\hspace*{-0.5cm}\times\exp\left[-a^2VW_7\tr^2 (U-U^{-1})-a^2 VW_8\tr (U^2+U^{-2})\right]{\det}^{\nu}U.\label{ZnuNf}
\end{eqnarray}
%%%%%%%%%%%%%%%%%%%%%%%%%%%%%%%%%%%%%%%%%%%%%%%%%%%%%%%%
Hereby we adopt the conventions of \cite{Damgaard:2010cz} for the 
LECs $W_{6/7/8}$. 
The partition function~\eqref{ZnuNf} is obtained in the $\epsilon$-regime of chiral perturbation theory (chPT) which is the limit of large volume, $V\to\infty$, where $\hat{m}=mV\Sigma$, $\hat{a}_{6/7}^2=-a^2VW_{6/7}$ and $\hat{a}_8^2=a^2VW_{8}$ are fixed\footnote{For $\hat{a}_{6/7/8}$ we employ the sign convention of \cite{Kieburg:2013xta}.}. In this limit there is an exact equivalence between the partition functions of chiral perturbation theory and chiral RMT \cite{Shuryak:1992pi,Verbaarschot:1994qf,Osborn:1998qb}.
The corresponding RMT is
\begin{eqnarray}\label{model}
 D_\W=\left(\begin{array}{cc} a A & W \\ -W^\dagger & a B \end{array}\right)+m_6\eins+\lambda_7\gamma_5,
\end{eqnarray}
where the random matrix $W$ generates the $a\to 0$ limit of Eq.~\eqref{ZnuNf}, the Hermitian matrices $A$ and $B$ generate the term proportional to $W_8$ and the two scalar random variables $m_6$ and $\lambda_7$ generate the terms proportional to $W_6$ and $W_7$, respectively. 
%%%%%%%%%%%%%%%%%%%%%%%%%%%%%%%%%%%%%%%%
All random variables are Gaussian distributed. In the microscopic limit, 
where  the matrix dimension $n\to\infty$ 
while fixing $\widehat{m}_6=2nm_6$, $\widehat{\lambda}_7=2n\lambda_7$ and $\widehat{a}_8^2=na^2/2$ the random matrix partition function reduces to the
chiral partition function~(\ref{ZnuNf}).

\begin{figure}[h!]
 \includegraphics[width=\textwidth,height=5cm]{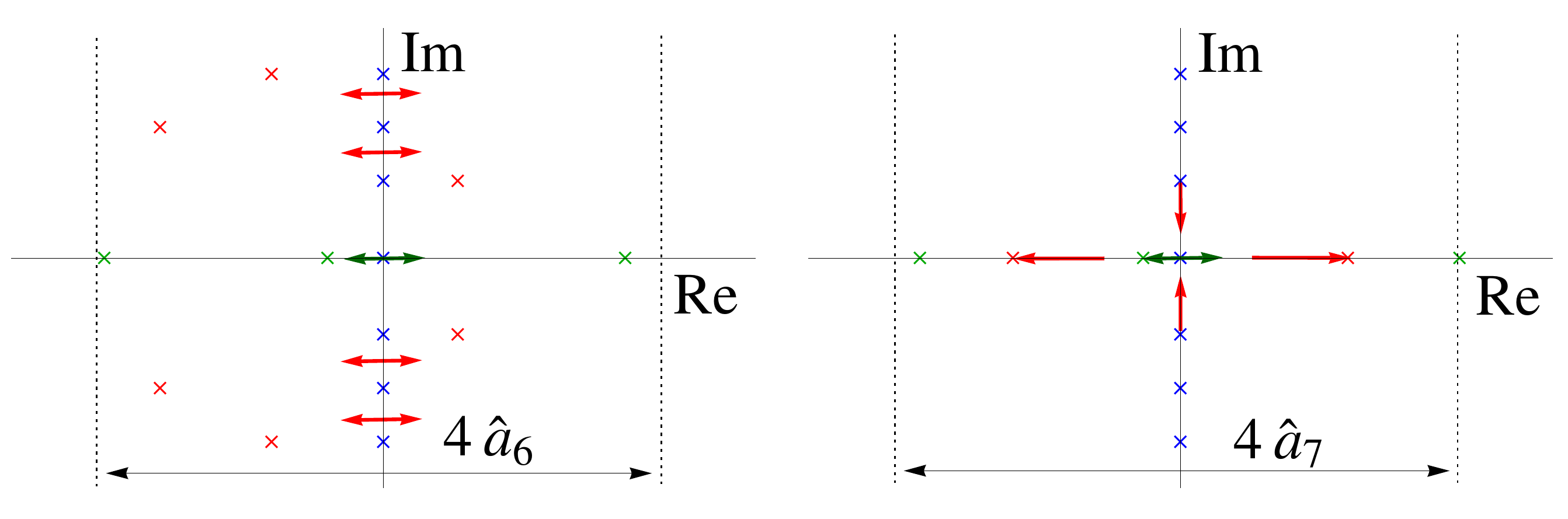}
 \caption{\label{fig1}Schematic plots of the effects of $W_6$ (left plot) and of $W_7$ (right plot) on the microscopic spectrum. The former zero modes (green crosses) are broadened by a Gaussian while the complex conjugate pairs (red crosses) are broadened parallel to the real axis by $W_6$ and are pushed into the real axis by $W_7$.
}
\end{figure}

First, let us focus on the effect of $W_6$ and $W_7$ and set $W_8=0$. 
In this case the massive Dirac operator reads
\be
D_\W= D_\W|_{a=0} +(m +m_6)\eins +\lambda_7 \gamma_5,\label{massDirac}
\ee
with eigenvalues given by
\be
z_\pm = m+m_6\pm \imath\sqrt{\lambda_\W^2-\lambda_7^2},
\label{ev67}
\ee
where we denote an eigenvalue of the continuum Dirac operator $D_\W|_{a=0}$ by $\imath\lambda_\W$.
In Fig.~\ref{fig1} the different effects of $W_{6/7}$ are shown 
schematically.

The LEC $W_6$ broadens the microscopic spectrum   
parallel to the real axis by a Gaussian with a width proportional to $\widehat{a}_6$,
while the shape of the spectrum projected to the real axis is not changed at all.  The reason is the additive nature of $\widehat{m}_6$ to the eigenvalues~\eqref{ev67} resulting in a convolution. The effect of $W_7$ is more drastic since the purely imaginary eigenvalues invade the real axis through the origin. 
Only the former zero modes are broadened by a Gaussian 
with width proportional to $\widehat{a}_7$.
% since then $\widehat{\lambda}_7$ is additive to the eigenvalues~\eqref{ev67}, too. 

\section{The eigenvalue densities and their dependence on $W_{6/7/8}$}\label{densities}

Due to $\gamma_5$-Hermiticity the eigenvalues of the
 Wilson Dirac operator come in complex conjugate pairs or are real. A Dirac operator with fixed index $\nu$ has $\nu$ 
generic real eigenvalues which correspond to the zero modes in 
the continuum limit. In addition,  other real 
modes resulting from complex conjugate pairs entering the real axis pairwise.
 Moreover, all real modes have non-zero chirality while the 
chirality of the complex modes vanishes. 
Thus, one can define three different eigenvalue densities, 
namely the one of the complex eigenvalues, $\rho_{\rm c}$, 
and the one of all real modes, $\rho_{\rm real}=\rho_{\rm right}+\rho_{\rm left}$,
 splitting into the density of the right handed real modes 
($\langle\psi|\gamma_5|\psi\rangle>0$), $\rho_{\rm right}$, 
and the one of the left handed real modes ($\langle\psi|\gamma_5|\psi\rangle<0$), $\rho_{\rm left}$. Another important distribution is the distribution of the chirality over the real eigenvalues  \cite{Akemann:2010zp}
\begin{equation}\label{chidef}
\rho^\nu_{\rm \chi}(\widehat{\lambda}) 
 \equiv  {\sum}_{\widehat{\lambda}_k \in {\mathbb R}}
\delta(\widehat \lambda -\widehat \lambda_k)  {\rm sign }\langle k| \gamma_5 |k \rangle,
\end{equation}
which is equal to the imaginary part of the resolvent
 \be
\label{chicond}
\rho^\nu_{\rm \chi}(\widehat{\lambda}) =\lim_{V\to\infty}\frac{1}{\pi}{\rm Im}
\left[G^\nu(\widehat{\lambda})\equiv \left\langle\tr\frac{1}{V\Sigma D_\W+\widehat{\lambda}\eins-\imath\epsilon\gamma_5}\right\rangle\right].
\ee
 With the help of this distribution one can define the density of additional real modes $\rho_{\rm add}=\rho_{\rm real}+\rho_\chi$. The splitting of the level density of real modes into $\rho_{\rm add}$ and $\rho_\chi$ has proven quite convenient \cite{Kieburg:2011uf,Kieburg:2013xta} as will be seen in the ensuing discussion. This discussion summarizes the extensive analytical study performed in \cite{Kieburg:2013xta}.

\begin{figure}[h!]
 \includegraphics[width=\textwidth,height=6cm]{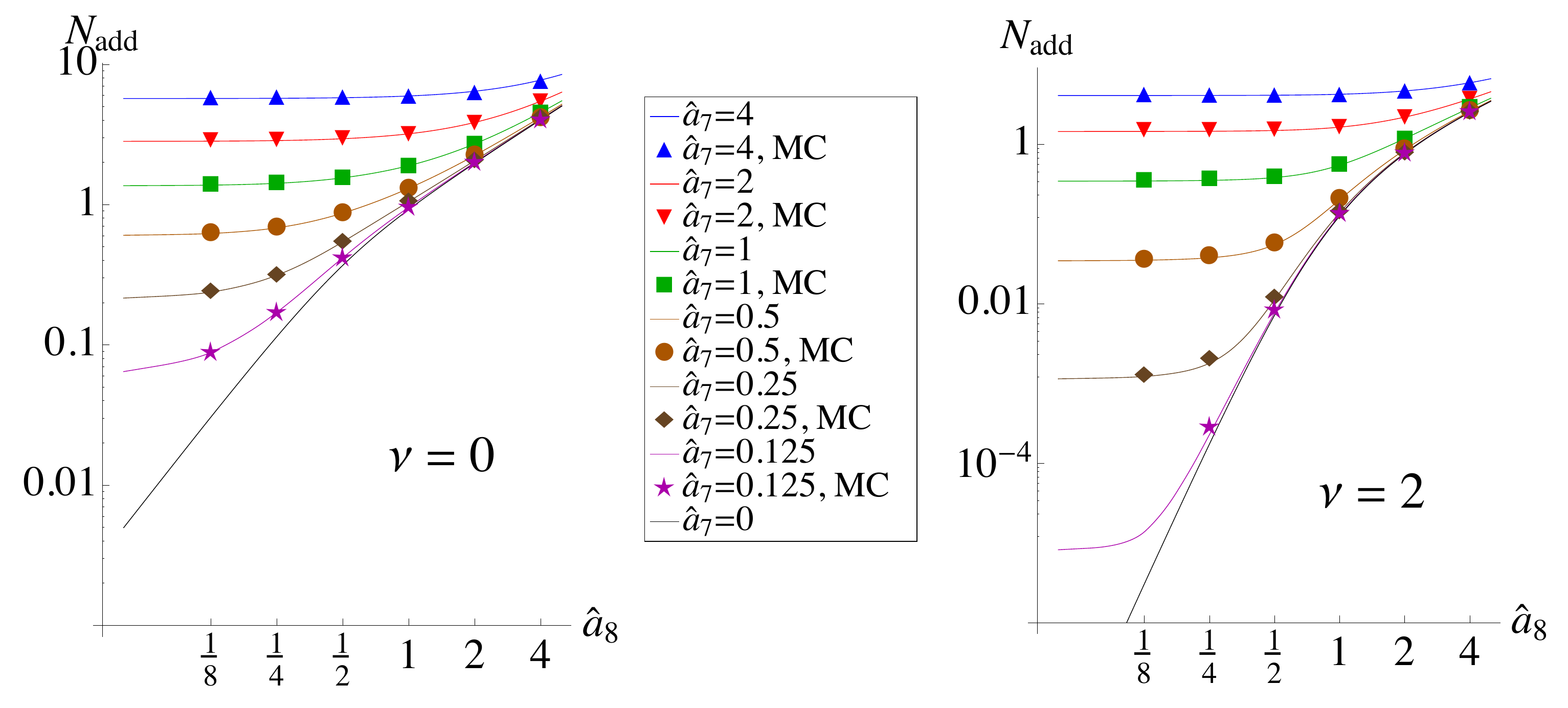}
 \caption{\label{fig3} Log-log plots of the additional real modes versus $\widehat{a}_8$ for $\nu=0$ (left plot) and $\nu=2$ (right plot). 
The analytical results (solid curves) are compared to Monte-Carlo 
simulations of RMT (symbols). 
Notice that a non-zero value of $\widehat{a}_7$ yields a saturation at small $\widehat{a}_8$.}
\end{figure}

First we discuss the additional real modes. The average number of additional real modes,
\begin{equation}\label{Nadd-def}
N_{\rm add}=\int_{-\infty}^\infty\rho_{\rm add}(\widehat{x})d\widehat{x},
\end{equation}
has shown to be a powerful quantity for measuring the 
strength of the lattice artifacts. Since we integrate 
over the real axis $N_{\rm add}$ is independent of $W_6$.  
At small lattice spacing $\widehat{a}_{7/8}\approx0.1$ 
the average number of additional real modes behaves as $\widehat{ a}^{2\nu+2}$ 
while it scales linearly with ${\widehat a}$ and becomes independent of $\nu$ in the limit of large lattice spacing, see Fig.~\ref{fig3}. Thus, at small lattice spacing almost all additional real modes come from the sector with index $\nu=0$ 
given by
\begin{eqnarray}
  N_{\rm add}^{\nu=0}&\overset{{\widehat{a}}\ll1}{=}&2Va^2(W_8-2W_7).\label{addrealsmall}
\end{eqnarray}
The distribution $\rho_{\rm add}$ lives on the scale $\widehat a$ at small lattice spacing and on the scale $\widehat a^2$ at large lattice spacing. The latter limit is close to the mean field limit where
the support of the real part of the eigenvalues scales with ${\widehat a}^2$.

\begin{figure}[h!]
 \includegraphics[width=7.5cm,height=6cm]{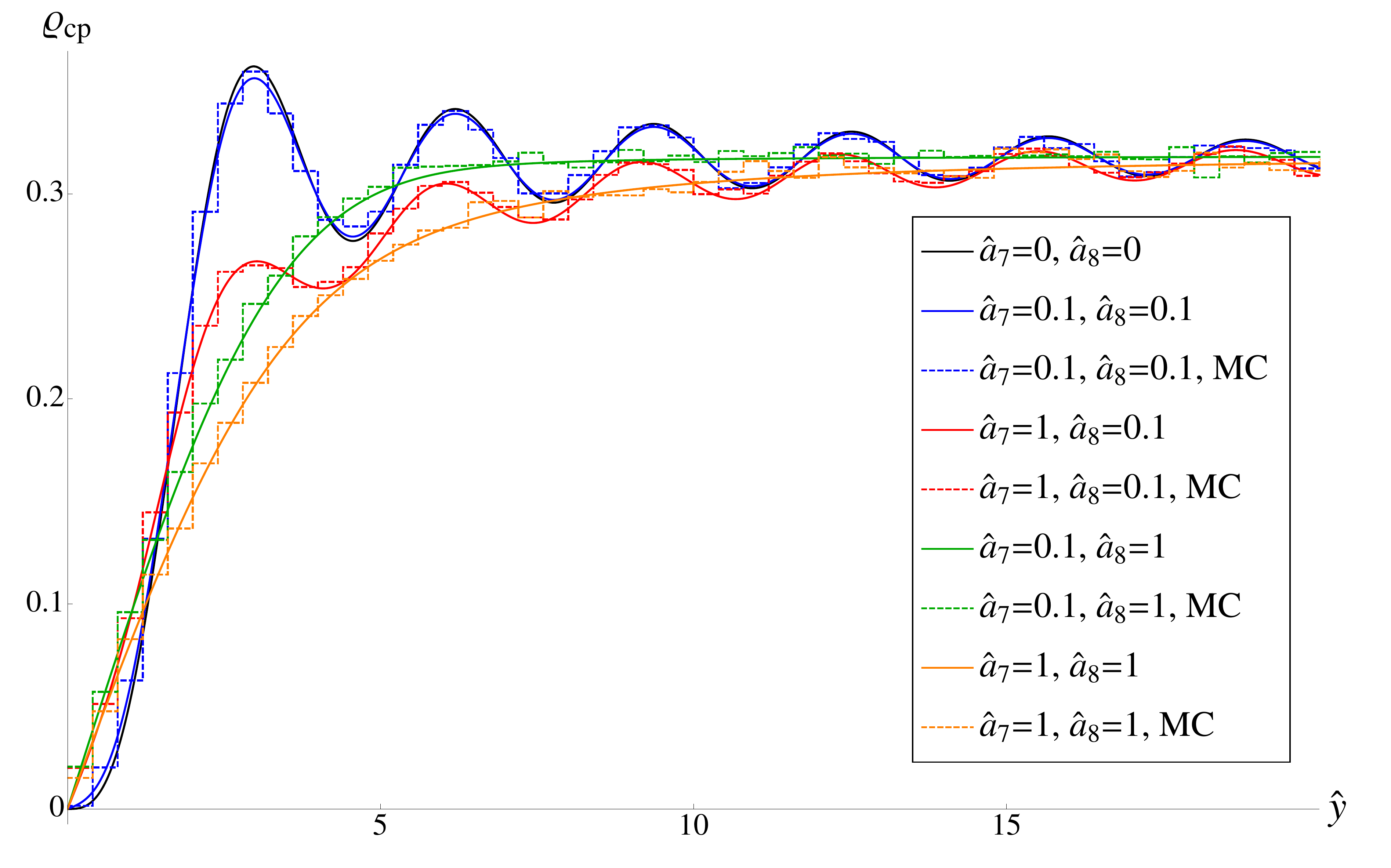}\hfill\includegraphics[width=7.5cm,height=6cm]{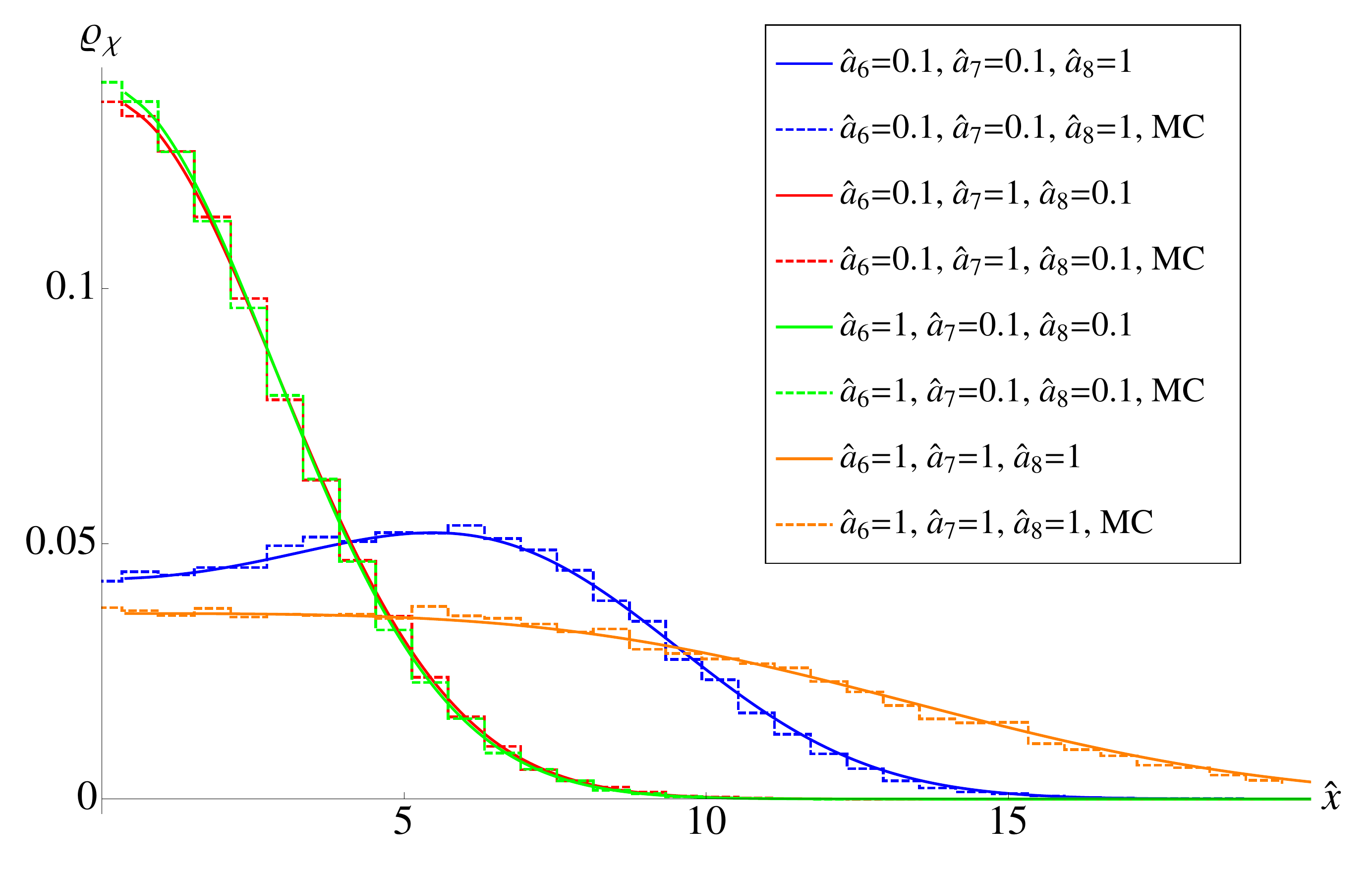}
 \caption{\label{fig4} Analytical results for the projected level density of complex eigenvalues onto the imaginary axis, $\rho_{\rm cp}$, (left)
 and for the distribution of the chirality over the real modes, $\rho_\chi$, (right) compared
to Monte-Carlo simulations of Wilson RMT for $\nu=1$. 
Notice that both distributions are symmetric around the origin 
in the quenched theory and we plotted only the positive imaginary and real axis, respectively. The black curve in the left plot is the continuum result and shows good agreement with $\rho_{\rm cp}$ for small 
lattice spacing while the deviation is quite substantial for large lattice spacing.}
\end{figure}
 
The behavior of the projected density of complex eigenvalues,
\begin{equation}\label{rhocp-def}
 \rho_{\rm cp}(\widehat{y})=\int_{-\infty}^\infty\rho_{\rm c}(\widehat{x}+\imath \widehat{y})d\widehat{x},
\end{equation}
when varying $W_7$ and $W_8$ can be seen in Fig.~\ref{fig4}. Also this quantity is $W_6$ independent since the integration cancels the convolution 
with the random variable $\widehat{m}_6$ and thus increases the statistics of the numerical simulation. The LEC $W_8$ smoothens
 the distribution such that the oscillations completely disappear 
when increasing this LEC. In contrast to this effect, $W_7$ 
dampens the height of $\rho_{\rm cp}$ near the real axis since 
the complex eigenvalue pairs are pushed into the real axis while 
the oscillations seem to be more persistent with regard to this 
LEC. Luckily the distribution $\rho_{\rm cp}$ smoothly 
converges to the continuum result in the limit of small lattice spacing 
and gives a hint that it is still a good quantity for extracting 
the chiral condensate via the Banks-Casher relation,
\be\label{Banks-Casher}
\Delta\overset{{\widehat{a}}\ll1}{=}\frac \pi{\Sigma V},
\ee
where $\Delta$ is the average level spacing of the imaginary part of the eigenvalues several eigenvalue spacings from the origin. Furthermore,
 the section of $\rho_{\rm c}$ parallel to the real axis is a Gaussian of width
\begin{eqnarray} \label{width}
 \frac{\sigma^2}{\Delta^2}&\overset{{\widehat{a}}\ll1}{=}&\frac 4{\pi^2}a^2 {V(W_8-2W_6)}
\end{eqnarray}
at small lattice spacing and becomes box-shaped on the scale $\widehat{a}_8^2=a^2VW_8$ in the limit of large lattice spacing as well as in the mean field limit.

Finally, we consider the distribution of the chirality over the real 
eigenvalues, $\rho_\chi$. This distribution is always normalized 
to  the index of the Dirac operator, i.e. 
$\int \rho_\chi(\widehat{x})d\widehat{x}=\nu$. 
At small lattice spacing the effects of $W_6$ and $W_7$ are almost the same and yield a Gaussian broadening on the scale $\widehat{a}$. In particular the variance of this distribution at fixed index $\nu$ takes a simple form, i.e.
\begin{eqnarray}\label{chismall}
\frac{\langle x^2\rangle_{\rho_\chi}}{\Delta^2}
&\overset{{\widehat{a}}\ll1}{=}&\frac 8 {\pi^2}Va^2(\nu W_8-W_6-W_7),\ \nu>0,
\end{eqnarray}
which is convenient for measuring the LECs. 
Moreover $\rho_\chi$ dominates the level density of the 
real eigenvalues at small lattice spacing, i.e. 
$\rho_\chi{\approx}\rho_{\rm real}$ (for $\widehat{a}\ll1$), since its height is of order
 ${\widehat a}^{-1}$ in contrast to the height of $\rho_{\rm add}$ which is of order ${\widehat a}^{2\nu+1}$.
 This behavior is
 reversed in the limit of large lattice spacing. 
Then the height of $\rho_\chi$ is of order ${\widehat  a}^{-2}$ with 
support of order ${\widehat a}^2$ while the height of 
$\rho_{\rm add}$ is of order ${\widehat a}^{-1}$. 
Hence one has 
$\rho_{\rm add}= \rho_{\rm real}+ \rho_{\chi} {\approx}\rho_{\rm real}$ (for $\widehat{a}\gg1$).

\section{Conclusions}\label{conc}

By utilizing powerful RMT techniques, see \cite{Kieburg,Kieburg:2011uf,Kieburg:2013xta},  we analytically calculated the spectral densities of the real and complex eigenvalues of the non-Hermitian Wilson Dirac operator $D_\W$ and summarized the most important results here. All the results were presented for the quenched theory but will be generalized to dynamical flavors in forthcoming publications.

We studied the explicit effects of all three LECs on the spectrum of the Wilson Dirac operator. Thereby we derived simple relations between the LECs, $\Sigma$ and $W_{6/7/8}$, and some measurable quantities like the average level spacing of the projected eigenvalues onto the imaginary axis and the average number of additional real modes, see Eqs.~\eqref{addrealsmall} and (\ref{Banks-Casher}-\ref{chismall}). Those relations apply at small lattice spacing $|a^2VW_{6/7/8}|\leq0.1$. In this regime these quantities may serve for fixing the LECs in lattice simulations. 
In particular the linear relations between $N_{\rm add}^{\nu=0}$, $\sigma$ and $\langle x^2\rangle_{\rho_\chi}$ are linearly
dependent requiring the consistency relation
\be
\frac{\langle  x^2 \rangle_{\rho_\chi}^{\nu = 1}}{\Delta^2} =
\frac{\sigma^2}{\Delta^2} + \frac 2{\pi^2} N^{\nu=0}_{\rm add}
\ee 
valid in the limit of small $a^2VW_{6/7/8}$. Such consistency relations as well as the relations ~(\ref{addrealsmall}) and~(\ref{Banks-Casher}-\ref{chismall}) will certainly improve the accuracy of the LECs fitted in \cite{DWW,DHS}.

Moreover we also considered the limit of large lattice spacing which is 
closely related to the mean field limit. The real part of the complex eigenvalues of $D_\W$ as well as the real eigenvalues themselves have a support
 in the interval $[-8 a^2VW_8,8 a^2VW_8]$ agreeing with the discussion about the complex eigenvalues in \cite{Kieburg:2012fw}.
 Although the impact of $W_6$ and $W_7$ on the density of the complex 
eigenvalues is negligible in the mean field limit, 
they have a strong impact on the distribution of the real eigenvalues.
In particular, in the mean field limit of the quenched theory, 
a non-zero $W_7$ generates a square root singularity 
at the boundary of the support of $\rho_{\rm add}$ while the density is uniform for  $W_7=0$.
 Nevertheless we expect that the phase diagram of the quenched as well as 
 the unquenched theory will not really change by this effect 
which will be checked  in a forthcoming publication.

\section*{Acknowledgement}

MK acknowledges partial financial support by the Alexander-von-Humboldt Foundation. JV and SZ acknowledge support by U.S. DOE Grant No. DE-FG-88ER40388. We thank Gernot Akemann, Poul Damgaard, Urs Heller and Kim Splittorff for fruitful discussions.

%%%%%%%%%%%%%%%%%%%%%%%%%%%%%%%%%%%%%%%%%%%%%%%%%%%%%%%%%%%%%%%%%%%%

\end{document}